\documentclass[10pt,conference]{IEEEtran}
\usepackage{graphicx} % Add all your packages here

\begin{document}
%\input{epsf}

\title{Fast Min-Sum Algorithms \\for Decoding of LDPC over $GF(q)$}

\author{\authorblockN{Xiaofei Huang, Suquan Ding, Zhixing Yang, Youshou Wu} \\
\authorblockA{School of Information Science and Technology, Tsinghua Univeristy, Beijing, P.~R.~China, 100084 \\
Email: huangxiaofei@ieee.org\\
(Accepted by IEEE Information Theory Workshop, Chengdu, China, 2006)}
}
% avoiding spaces at the end of the author lines is not a problem with
% conference papers because we don't use \thanks or \IEEEmembership
%Homer Simpson\authorrefmark{2},
%Montgomery Scott\authorrefmark{3} and
%Eldon Tyrell\authorrefmark{4}}
%Telephone: (800) 555--1212, Fax: (888) 555--1212}
%\authorblockA{\authorrefmark{4}Tyrell Inc., 123 Replicant Street, Los Angeles, California 90210--4321}}

\maketitle

\begin{abstract}
In this paper, we present a fast min-sum algorithm 
	for decoding LDPC codes over $GF(q)$.
Our algorithm is different from the one presented by David Declercq and Marc Fossorier in~\cite{DeclercqISIT05}
	only at the way of speeding up the horizontal scan in the min-sum algorithm.
The Declercq and Fossorier's algorithm speeds up the computation 
	by reducing the number of configurations, 
	while our algorithm uses the dynamic programming instead.
Compared with the configuration reduction algorithm,
	the dynamic programming one is simpler at the design stage 
	because it has less parameters to tune.
Furthermore, it does not have the performance degradation problem 
	caused by the configuration reduction because it searches the whole configuration space efficiently through dynamic programming.
Both algorithms have the same level of complexity
	and use simple operations which are suitable for hardware implementations.
\end{abstract}

\section{Introduction}

LDPC (low density parity check) codes are the state of art technology~\cite{Gallager:LDPCC:thesis,MacKay:GCBOVSM}
	for their near Shannon limit performance for channel error correction~\cite{Richardson:DOCAILDPCC}.
China has considered it for broadcasting digital video for terrestrial televisions.
Europe has accepted for its next generation broadcasting digital video using satellites (DVB-S2).
LDPC codes have also accepted or considered by many industry standards 
	such as IEEE 802.16 and IEEE 802.11n.
The LDPC codes defined over Galois field $GF(q)$ of order $q > 2$ have shown significant improved performance
	over binary LDPC codes.
	
David Declercq and Marc Fossorier presented in \cite{DeclercqISIT05} a fast min-sum algorithm
	for decoding LDPC codes over $GF(q)$.
It is a generalization of the normalized/offset min-sum algorithm from the Galois field $GF(2)$~\cite{Gallager:LDPCC:thesis,MacKay:GCBOVSM}
	to any Galois field, $GF(q)$ for any $q \ge 2$.
The Declercq and Fossorier's algorithm has much less complexity 
	than another generalization of the min-sum algorithm given in~\cite{Wymeersch2004}.
Their algorithm speeds up the computation 
	by reducing the number of configurations evaluated at the horizontal scan of the min-sum algorithm.

In this paper, we present another min-sum algorithm
	which different from the Declercq and Fossorier's one only at the horizontal scan.
We use the dynamic programming technique to speed up the horizontal scan 
	instead of reducing the number of configurations.
Both techniques have the same level of complexity.
The latter finds approximate solutions at the horizontal scan which may introduce some performance degradation,
	while the former finds exact solutions and does not cause any performance degradation.
The former is also simpler to design because it does not need to tune the balance between the configuration reduction
	and performance.

%%%%%%%%%%%%%%%%%%%%%%%%%%%%%%%%%%%%%%%%%%%%%%%%%%%%%%%%%%%%%%%%%%%%%%%%%%%%%%%%%%%%%
\section{Generalized Min-Sum Algorithms for Decoding LDPC Codes over $GF(q)$}
%%%%%%%%%%%%%%%%%%%%%%%%%%%%%%%%%%%%%%%%%%%%%%%%%%%%%%%%%%%%%%%%%%%%%%%%%%%%%%%%%%%%%

\subsection{The Problem Statement}

LDPC codes belong to a special class of linear block codes
	whose parity check matrix $H$ has a low density of ones.
For a LDPC code over $GF(q)$, its parity check matrix $H$ has elements $h_{mn}$ defined over $GF(q)$, $h_{mn} \in GF(q)$.
Let the code word length be $N$ (the number of symbols),
then $H$ is a $M \times N$ matrix, where $M$ is the number of rows. 
Each row of $H$ introduces one parity check constraint on input data $x=(x_1, x_2, \ldots, x_N)$, i.e.,
\[ \sum^{N}_{n=1} h_{mn} x_n = 0,~~\mbox{for $m=1,2,\ldots, M$} \ . \]
Putting the $m$ constraints together, we have $H x^T = 0$.

Let function $f_n (x_n)$ be defined as 
\[f_n (x_n) = - \ln p(x_n/y_n) \ , \]
where $p (x_n / y_n)$ is the conditional distribution of input data symbol $n$ at value $x_n$ 
	given the output data symbol $n$ at value $y_n$.
$f_n(0) - f_n(x_n)$, which is equal to $\ln (p(x_n/y_n) / p(0/y_n))$,
	is the log-likelihood ratio (LLR) of input data symbol $n$ at value $x_n$ versus value $0$.

In those notations, the maximum likelihood decoding can be formulated as a constrained optimization problem,
\begin{equation}
\min_{x_1, x_2, \ldots, x_N} \sum^n_{n=1} f_n (x_n) \quad \mbox{ s.t.~(subject to) $Hx^T = 0$} \ . 
\label{original_problem}
\end{equation}
The above function is called the objective function for decoding a LDPC code.
The decoding problem is, thus, transferred as finding the global minimum of a multi-variate objective function.

Let $X$ be the set of all variables.
Given the $m$th constraint be $H_m x^T = 0$,
	let $X_m$ be the subset of variables corresponding to the non-zero elements in $H_m$, 
	i.e.,
\[ X_m \equiv \{x_n|h_{mn} \not = 0\} \ . \]
Let $f_{X_m}(X_m)$ be a function defined over $X_m$ as
\[ f_{X_m}(X_m) = \left\{ \begin{array}{ll}
                      0, & \mbox{if $H_m x^T = 0$}; \\
                      \infty, & \mbox{ otherwise}.
                     \end{array}
             \right. \]
$f_{X_m}(X_m)$ is called the constraint function representing the $m$th constraint.
Using the constraint functions, 
	the decoding problem~(\ref{original_problem}) can be reformulated as a unconstrained combinatorial optimization problem
	of the following objective function,
\begin{equation}
E(x) = \sum^{M}_{m=1} f_{X_m}(X_m) + \sum^N_{n=1} f_n (x_n) \ . 
\label{objective_function_LDPC}
\end{equation}

\subsection{Generalized Min-Sum Algorithm for LDPC over $GF(q)$}

Dr. Wiberg~\cite{Wiberg:thesis} developed the min-sum algorithm as a generalization of the Viterbi algorithm.
The min-sum algorithm is also proposed in~\cite{Fossorier99}
	as an approximation to the belief propagation (BP) algorithm~\cite{Pearl88,Kschischang01}.
It is also referred to as the BP-based algorithm.
The min-sum algorithm is a soft-decision, iterative algorithm for decoding binary-LDPC codes.

Conventionally, a LDPC code is represented as a Tanner graph,
	a graphical model useful at understanding code structures and decoding algorithms.
A Tanner graph is a bipartite graph with variable nodes on one side
	and constraint nodes on the other side. 
Edges in the graph connect constraint nodes to variable nodes. 
A constraint node connects to those variable nodes that are contained in the constraint.
A variable node connects to those constraint nodes that use the variable in the constraints.
Constraint nodes are also referred to as check nodes.
During each iteration of the min-sum algorithm,
	messages are flowed from variables nodes to the check nodes first,
	then back to variable nodes from check nodes.

Let ${\cal N}(m)$
	be the set of variable nodes that are connected to the check node $m$.
Let $ {\cal M}(n)$ be the set of check nodes that are connected to the variable node $n$.
Let symbol `$\setminus$' denotes the set minus.
${\cal N}(m) \setminus n$ denotes the set of variable nodes excluding node $n$ 
	that are connected to the check node $m$.
${\cal M}(n) \setminus m$ stands for the set of check nodes excluding the check node $m$ 
	which are connected to the variable node $n$.

The generalization of the min-sum algorithm for decoding LDPC codes over $GF(q)$
	is straightforward.
At iteration $k$, 
	let $Z^{(k)}_{mn}(x_n)$ denote the message sent from variable node $n$ to check node $m$.
$Z^{(k)}_{mn}(0) - Z^{(k)}_{mn}(x_n)$ is the log-likelihood ratio (LLR)
	of the $n$-th input symbol having the value $x_n$ versus $0$, 
	given the information obtained via the check nodes other than check node $m$.
Let $L^{(k)}_{mn}(x_n)$ denote the message sent from check node $m$ to variable node $n$.
$L^{(k)}_{mn}(x_n)$ is the log-likelihood ratio that the check node $m$ is satisfied 
	when input symbol $n$ is fixed to value $0$ versus value $x_n$
	and the other symbols are independent with log-likelihood ratios,
\[ Z_{mn^{'}}(0)  - Z_{mn^{'}}(x_n^{'}), \quad n^{'} \in {\cal N}(m) \setminus n. \ . \]

The pseudo-code of the generalized min-sum algorithm for decoding LDPC over $GF(q)$ is given as follows.

{\em Initialization} 

For $n = 1, 2, \ldots, N$, and $m = 1, 2, \ldots, M$,
~~~~\[ Z^{(0)}_{mn} (x_n) =  f_n (x_n) \ . \]

{\em Iteration (k = 1, 2, 3, \ldots})
\begin{enumerate}
\item {\em Horizontal scan}

{\em Compute $L^{(k)}_{mn}(x_n)$, for each $x_n \in GF(q)$, }
\begin{eqnarray}
L^{(k)}_{mn} (x_n) &=& \min_{X_m \setminus x_n} 
	\sum_{n^{'} \in {\cal N}(m) \setminus n} Z^{(k-1)}_{mn^{'}} (x_{n^{'}}) \label{horizontal_scan} \\
	\mbox{s.t. }&& \sum_{n^{'} \in {\cal N}(m)} h_{mn^{'}} x_{n^{'}} = 0 \ ,  \nonumber
\end{eqnarray}

{\em Normalize $L^{(k)}_{mn}(x_n)$}

For each $m$, and each $n \in {\cal N}(m)$, offsetting $L^{(k)}_{mn}(x_n)$ by $L^{(k)}_{mn}(0)$,
\[ L^{(k)}_{mn}(x_n) \Leftarrow L^{(k)}_{mn}(x_n) - L^{(k)}_{mn}(0) \ . \]

\item {\em Vertical scan}

For $n = 1, 2, \ldots, N$,
\begin{equation}
Z^{(k)}_{mn}(x_n) = f_{n} (x_n) + \sum_{m^{'} \in {\cal M}(n) \setminus  m} L^{(k)}_{m^{'}n}(x_n)  \ . 
\label{vertical_scan1}
\end{equation}
\item {\em Decoding}

For each symbol, compute its posteriori log-likelihood ratio (LLR)
\begin{equation}
Z^{(k)}_n(x_n) = f_n(x_n) + \sum_{m \in {\cal M}(n)} L^{(k)}_{mn}(x_n) \ . 
\label{vertical_scan2}
\end{equation}
Then estimate the original codeword ${\hat x}^{(k)}$,
\[
{\hat x}^{(k)}_n = \arg \min_{x_n} Z^{(k)}_n(x_n), \quad \mbox{for $n = 1, 2, \ldots, N$} \ .
\]
If $H~({\hat x}^{(k)})^T = 0$ or the iteration number exceeds some cap, 
	stop the iteration and output ${\hat x}^{(k)}$ as the decoded codeword.
\end{enumerate}

In the above algorithm,
	$Z^{(k)}_{n}(0) - Z^{(k)}_n(x_n)$ is the posteriori LLR for variable $x_n$ at iteration $k$.

One way to possibly improve the performance of the generalized min-sum algorithm
	is to modify the Eq.~(\ref{vertical_scan1}) and Eq.~(\ref{vertical_scan2}) as
\[ Z^{(k)}_{mn}(x_n) = f_{n} (x_n) + \alpha_{k} \sum_{m^{'} \in {\cal M}(n) \setminus  m} L^{(k)}_{m^{'}n}(x_n)  \ , \]
\[ Z^{(k)}_n(x_n) = f_n(x_n) + \alpha_{k} \sum_{m \in {\cal M}(n)} L^{(k)}_{mn}(x_n) \ , \]
where $\alpha_k$ is a scaling constant at iteration $k$ satisfying $0 < \alpha_k < 1$.
With these modifications, the decoding algorithm is called the normalized min-sum algorithm.

Another way to possibly improve the performance 
	is to modify the Eq.~(\ref{vertical_scan1}) and Eq.~(\ref{vertical_scan2}) as
\[ Z^{(k)}_{mn}(x_n) = f_{n} (x_n) + \sum_{m^{'} \in {\cal M}(n) \setminus  m} \max(L^{(k)}_{m^{'}n}(x_n) - \beta_k, 0)  \ , \]
\[ Z^{(k)}_n(x_n) = f_n(x_n) + \sum_{m \in {\cal M}(n)} \max(L^{(k)}_{mn}(x_n) - \beta_k, 0) \ , \]
where $\beta_k$ is an offset constant at iteration $k$ satisfying $\beta_k > 0$.
With these modifications, the decoding algorithm is called the offset min-sum algorithm.

To possibly maximize the decoding power, 
	the scaling factor $\alpha_k$ or the offset constant $\beta_k$ 
	can be determined through experiments or the density evolution method~\cite{JinghuChenThesis}.

%%%%%%%%%%%%%%%%%%%%%%%%%%%%%%%%%%%%%%%%%%%%%%%%%%%%%%%%%%%%%%%%%%%%%%%%%%%%%%%%%%%%%%%
\subsection{Horizontal Scan via Dynamic Programming}
%%%%%%%%%%%%%%%%%%%%%%%%%%%%%%%%%%%%%%%%%%%%%%%%%%%%%%%%%%%%%%%%%%%%%%%%%%%%%%%%%%%%%%%
\label{section_dynamic_programming}

Our algorithm for the horizontal scan is based on dynamic programming~\cite{Coffman76}, 
	which is, in principle, similar to the Viterbi algorithm~\cite{forney73}
	for decoding convolutional codes.
It is a linear complexity algorithm for the minimization problem defined in (\ref{horizontal_scan}) 
	as long as all $x_n$s are in finite domains.
For decoding LDPC codes over $GF(q)$, all variables are defined in $GF(q)$, a finite domain.
The algorithm is applicable for this special case.

With loss of generality, 
	we explain the dynamic programming algorithm with the assumption of $h_{mn} \not = 0$ for all $n$s
	to simplify notations.
When some of elements $h_{mn}$s are zero, 
	we can apply the same algorithm simply on those variables 
	with non-zero coefficients $h_{mn}$ in the parity check constraint $\sum_n h_{mn} x_n = 0$.
	
To simplify notations further,
	we define $g_n(x_n)$ as
\[ g_n (x_n) = Z^{(k-1)}_{mn} (x_{n}) \ . \]
The minimization problem defined in (\ref{horizontal_scan}) can be rewritten in a more succinct form
\begin{equation}
L^{(k)}_{mn}(x_n) = \min_{X \setminus x_n} \sum^N_{n^{'}=1} g_{n^{'}}(x_{n^{'}}), \mbox{ s.t.} \sum^N_{n^{'}=1} h_{mn^{'}} x_{n^{'}} = 0.  \label{cooperative_optimization3d}
\end{equation}
%\begin{eqnarray}
%L^{(k)}_{mn}(x_n) & = & \min_{X \setminus x_n} 
%	\sum^N_{n^{'}=1} g_{n^{'}}(x_{n^{'}}) \label{cooperative_optimization3d} \\
%	& \mbox{s.t.} & \sum^N_{n^{'}=1} h_{mn^{'}} x_{n^{'}} = 0 \ . \nonumber 
%\end{eqnarray}

We need to solve the minimization problem (\ref{cooperative_optimization3d}) $N$ times,
	one for each $x_n$, $n=1, 2, \ldots, N$.
This task can be done through two scans, each scan defines a dynamic programming process.
One scan is started from variable $x_1$ and ended at variable $x_N$, so called the left scan.
The other has the reverse order, from $x_N$ to $x_1$, so called the right scan.

Each scan has $N-1$ steps, step $n = 1, 2, \ldots, N-1$. 
We take the left scan as the case study.
The right scan can be derived simply by reversing the order of variables.

For the left scan, at step $n$, we use variable $s_n$, $s_n \in GF(q)$,
	to represent the result of the following summation,
\[ s_n = \sum^n_{n^{'}=1} h_{mn^{'}} x_{n^{'}} \ . \]
Also, we assign a real value $r^{L}_n (s_n)$ for each state $s_n$, 
	which stores the result of the following constrained optimization problem,
\[ r^{L}_n (s_n) = \min_{x_1, \ldots, x_n} \sum^n_{n^{'} = 1 } g_{n^{'}}(x_{n^{'}})
	\quad \mbox{s.t. }\sum^n_{n^{'}=1} h_{mn^{'}} x_{n^{'}} = s_n \ , \]
where the superscript "L" stands for the left scan.

When $n=1$, $r^{L}_1(s_1)$ is initialized as
\[ r^{L}_1 (s_1) = g_1 (h^{-1}_{m1}s_1) \ . \]
At each step $n$, $n = 2, 3, \ldots, N-1$,
	the dynamic programming computes $r^{L}_n (s_n)$ for each state $s_n$, $s_n \in GF(q)$, as follows
\begin{eqnarray}
r^{L}_n (s_n) &=& \min_{x_n, s_{n-1}} g_n(x_n) + r^{L}_{n-1} (s_{n-1}) \\
&\mbox{s.t.} & \quad s_{n-1} + h_{mn} x_n = s_n \ , \nonumber
\label{dp1}
\end{eqnarray}

Similarly, for the right scan, 
	when $n=N$, $r^{R}_1(s_1)$ is initialized as
\[ r^{R}_N (s_N) = g_N (h^{-1}_{m,N}s_N) \ . \]
We compute $r^{R}_n (s_n)$, for $n = N-1, N-2, \ldots, 2$, as follows
\begin{eqnarray}
r^{R}_n (s_n) &=& \min_{x_n, s_{n+1}} g_n(x_n) + r^{R}_{n+1} (s_{n+1}) \\
&\mbox{s.t.}& \quad s_{n+1} + h_{mn} x_n = s_n \ , \nonumber
\label{dp2}
\end{eqnarray}

We can obtain the minimization result for (\ref{cooperative_optimization3d}) from $r^{L}_n (s_n)$ and $r^{R}_n (s_n)$ directly.
For $1 < n < N$, we have
\begin{eqnarray}
L^{(k)}_{mn}(x_n)&=& \min_{s_{n-1}, s_{n+1}} r^{L}_{n-1} (s_{n-1}) + r^{R}_{n+1} (s_{n+1}) \label{dp3a} \\
&\mbox{s.t.}&  s_{n-1} + h_{mn}x_{n} + s_{n+1} = 0 \ . \nonumber
\end{eqnarray}	
We can rewrite Eq.~(\ref{dp3a}) to have a form clearer for computing
\begin{equation}
L^{(k)}_{mn}(x_n)= \min_{s_{n-1}} r^{L}_{n-1} (s_{n-1}) + r^{R}_{n+1} (-(s_{n-1} + h_{mn}x_{n})) \ .
\label{dp3e}
\end{equation}
For $GF(q),q = 2^m$, Eq.~(\ref{dp3e}) can be simplified further to
\begin{equation}
L^{(k)}_{mn}(x_n)= \min_{s_{n-1}} r^{L}_{n-1} (s_{n-1}) + r^{R}_{n+1} (s_{n-1} + h_{mn}x_{n}) \ .
\label{dp3f}
\end{equation}

When $n = N$, the result is
\begin{equation}
L^{(k)}_{m,N}(x_N) = r^{L}_{N-1} (-h_{mN}x_{N}) \ .
\label{dp3b}
\end{equation}	
When $n = 1$, the result is
\begin{equation}
L^{(k)}_{m,1}(x_1) = r^{R}_{2} (-h_{m1}x_{1}) \ .
\label{dp3c}
\end{equation}	

\subsection{Computational Complexity}

At each iteration, the vertical scan of the generalized (normalized/offset) min-sum algorithm 
	has the computational complexity of ${\cal O}(N d_v q)$,
	where $d_v$ is the average variable degrees.
	
For each constraint (check node), the dynamic programming horizontal scan 
	carries ${\cal O}(d_c(m) q^2)$ minimization operations and the same number of addition operations.
$d_c(m)$ here is the degree of the $m$th check node.
In total, the complexity of the horizontal scan is ${\cal O}(M d_c q^2)$,
	where $d_c$ is the average check node degrees.

If we reduce the number of candidate symbols for each variable from $q$ to $n_m$ best candidate symbols,
	the complexity of the horizontal scan is ${\cal O}(M d_c n_m q)$.
The complexity of the horizontal scan of the algorithm proposed in \cite{Wymeersch2004} is ${\cal O}(M d_c q^2)$.
If $n_m$ is small compared to $q$, 
	the complexity of our algorithm can be remarkably reduced.
For example, for codes over $GF(256)$, we reduced $q=256$ to $n_m=16$ without noticing much degradation 
	in performance in our experiments.
In this case, the complexity of the dynamic programming horizontal scan is reduced by factor $16$.
Nevertheless, such a speedup can cause degradation in performance of the decoding algorithm
	if $n_m$ is too small compared with $q$.
Furthermore, the degree of the degradation could vary from one code structure to another code structure.

%%%%%%%%%%%%%%%%%%%%%%%%%%%%%%%%%%%%%%%%%%%%%%%
\section{Experimental Results}
%%%%%%%%%%%%%%%%%%%%%%%%%%%%%%%%%%%%%%%%%%%%%%%
We have used the two LDPC codes offered by Davey and Mackay in \cite{DaveyMackay1998}
	to evaluate the performance of the generalized min-sum algorithm with the dynamic programming horizontal scan.
The first code is defined over $GF(4)$ of a code length $9,000$ and the second is over $GF(8)$ of a code length $6,000$.
The code rates of the both codes are $1/3$.

\begin{figure}
\centering
\includegraphics[width=8.0cm]{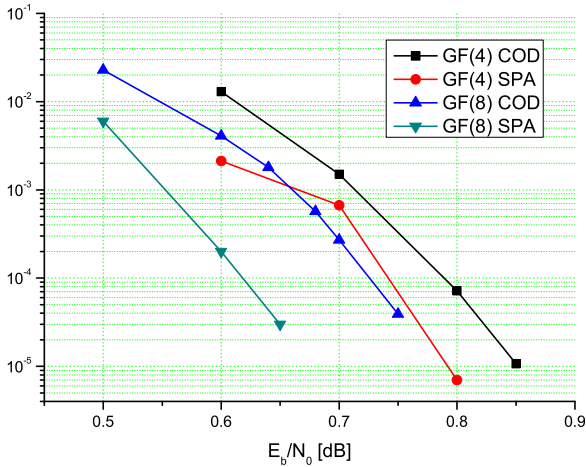}
\caption{The bit error rates (BER) of decoding a $GF(4)$ LDPC code 
	and a $GF(8)$ LDPC code using the standard sum-product algorithm (SPA) and our algorithm (COD).}
\label{DecodingLDPC}
\end{figure}

In our simulation, we use BPSK modulation and AWGN (additive White Gaussian Noise) channel.
Figure~\ref{DecodingLDPC} shows the performances of the normalized min-sum algorithm (COD) with the dynamic programming horizontal scan
	and the sum-product algorithm (SPA) at decoding both the codes.
The factor $\alpha$ used by the normalized min-sum algorithm is $0.865$ for the code over $GF(4)$ 
	and $0.820$ for the code over $GF(8)$.
%The belief propagation algorithm is also referred to as the sum-product algorithm in some literatures.
The normalized min-sum algorithm is a special case of a newly discovered optimization method called 
	the cooperative optimization (see \cite{HuangBookCCO,HuangISIT05}).
The maximum numbers of iterations for both the algorithms are all set to $300$.

From the figure we can see that 
	the performances of the normalized min-sum algorithm
	are very close to those of the sum-product algorithm.
The former is only around $0.05~dB$ away from the latter at decoding the $GF(4)$ LDPC code.
The degradation increases to $0.1~dB$ for the $GF(8)$ LDPC code which is still negligible.
The (normalized/offset) min-sum algorithm uses only additions and minimizations in its computation.
The SPA in its computation uses addition operations and expensive multiplication operations. 
The SPA in the log domain is a little bit more complex in computation than the COD for software implementations
	due to the table looking up operations, which are expensive for parallel hardware implementations.
Furthermore, the min-sum algorithm does not dependent on the channel estimate
	while the sum-product algorithm needs to estimate the variance of the channel noise.
The inaccuracy in the channel estimate can lead to noticeable performance degradations of the sum-product algorithm.

%%%%%%%%%%%%%%%%%%%%%%%%%%%%%%%%%%%%%%%%%%%%%%%
\section{Conclusion}
%%%%%%%%%%%%%%%%%%%%%%%%%%%%%%%%%%%%%%%%%%%%%%%

We have presented in this paper a general (normalized/offset) min-sum algorithm 
	for decoding LDPC codes over any Galois field $GF(q)$, $q \ge 2$.
To speed up the horizontal scan of the algorithm,
	the dynamic programming technique has been applied.
At each iteration, the computational complexity of the vertical scan of the algorithm 
	is ${\cal O}(N d_v q)$ and the computational complexity of the horizontal scan 
	is ${\cal O}(M d_v n_m q)$, $n_m < q$.
In our experiments, compared with the belief propagation algorithm,
	the generalized min-sum algorithm with the dynamic programming horizontal scan
	has only around $0.1~dB$ degradation in performance at water fall regions.
It is suitable for hardware implementations because it is simple in computation 
	and uses only minimization and addition operations.

\nocite{DaveyMackay1998,HuangISIT05,DeclercqISIT05,Barnault2003,Wymeersch2004,JinghuChenThesis,Gallager:LDPCC:thesis}
\bibliographystyle{../bib/IEEEtran}
%\bibliography{../bib/AIsfs}

\begin{thebibliography}{10}
\providecommand{\url}[1]{#1}
\def\UrlFont{\rmfamily}
\providecommand{\newblock}{\relax}
\providecommand{\bibinfo}[2]{#2}
\providecommand\BIBentrySTDinterwordspacing{\spaceskip=0pt\relax}
\providecommand\BIBentryALTinterwordstretchfactor{4}
\providecommand\BIBentryALTinterwordspacing{\spaceskip=\fontdimen2\font plus
\BIBentryALTinterwordstretchfactor\fontdimen3\font minus
  \fontdimen4\font\relax}
\providecommand\BIBforeignlanguage[2]{{%
\expandafter\ifx\csname l@#1\endcsname\relax
\typeout{** WARNING: IEEEtran.bst: No hyphenation pattern has been}%
\typeout{** loaded for the language `#1'. Using the pattern for}%
\typeout{** the default language instead.}%
\else
\language=\csname l@#1\endcsname
\fi
#2}}

\bibitem{DeclercqISIT05}
D.~Declercq and M.~Fossorier, ``Extended minsum algorithm for decoding {LDPC}
  codes over ${GF}(q)$,'' in \emph{Proceedings of IEEE International Symposium
  on Information Theory (ISIT)}, 2005, pp. 464--468.

\bibitem{Gallager:LDPCC:thesis}
R.~G. Gallager, ``Low-density parity-check codes,'' Ph.D. dissertation,
  Department of Electrical Engineering, M.I.T., Cambridge, Mass., July 1963.

\bibitem{MacKay:GCBOVSM}
D.~J.~C. MacKay and R.~M. Neal, ``Good codes based on very sparse matrices,''
  in \emph{Cryptography and Coding, 5th IMA Conference}, December 1995.

\bibitem{Richardson:DOCAILDPCC}
T.~J. Richardson, M.~A. Shokrollahi, and R.~L. Urbanke, ``Design of
  capacity-approaching irregular low-density parity-check codes,'' \emph{{IEEE}
  Transactions on Information Theory}, vol.~47, no.~2, pp. 619--637, February
  2001.

\bibitem{Wymeersch2004}
H.~Wymeersch, H.~Steendam, and M.~Moeneclaey, ``Log-domain decoding of {LDPC}
  codes over ${GF}(2^q)$,'' in \emph{The Proc. IEEE Intern. Conf. on Commun.},
  2004, pp. 772--776.

\bibitem{Wiberg:thesis}
N.~Wiberg, ``Codes and decoding on general graphs,'' Ph.D. dissertation,
  Department of Electrical Engineering, Linkoping University, Linkoping,
  Sweden, 1996.

\bibitem{Fossorier99}
M.~Fossorier, M.~Mihaljevic, and H.~Imai, ``Reduced complexity iterative
  decoding of low density parity check codes based on belief propagation,''
  \emph{IEEE Transactions on Communications}, vol.~47, pp. 673--680, May 1999.

\bibitem{Pearl88}
J.~Pearl, \emph{Probabilistic Reasoning in Intelligent Systems: Networks of
  Plausible Inference.}\hskip 1em plus 0.5em minus 0.4em\relax Morgan Kaufmann,
  1988.

\bibitem{Kschischang01}
F.~R. Kschischang, B.~J. Frey, and H.~andrea Loeliger, ``Factor graphs and the
  sum-product algorithm,'' \emph{IEEE Transactions on Information Theory},
  vol.~47, no.~2, pp. 498--519, February 2001.

\bibitem{JinghuChenThesis}
J.~Chen, ``Reduced complexity decoding algorithms for low-density parity check
  codes and turbo codes,'' Ph.D. dissertation, University of Hawaii, Dept. of
  Electrical Engineering, 2003.

\bibitem{Coffman76}
E.~G.~C. Jr., Ed., \emph{Computer and Job-Shop Scheduling}.\hskip 1em plus
  0.5em minus 0.4em\relax New York: Wiley-Interscience, 1976.

\bibitem{forney73}
J.~G.~D.~Forney, ``The {Viterbi} algorithm,'' \emph{Proc. IEEE}, vol.~61, pp.
  268--78, Mar. 1973.

\bibitem{DaveyMackay1998}
M.~C. Davey and D.~J.~C. MacKay, ``Low density parity check codes over
  {GF}(q),'' \emph{IEEE Communications Letters}, vol.~2, no.~6, pp. 165--167,
  June 1998.

\bibitem{HuangBookCCO}
X.~Huang, ``Cooperative optimization for solving large scale combinatorial
  problems,'' in \emph{Theory and Algorithms for Cooperative Systems}, ser.
  Series on Computers and Operations Research.\hskip 1em plus 0.5em minus
  0.4em\relax World Scientific, 2004, pp. 117--156.

\bibitem{HuangISIT05}
------, ``Near perfect decoding of {LDPC} codes,'' in \emph{Proceedings of IEEE
  International Symposium on Information Theory (ISIT)}, 2005, pp. 302--306.

\bibitem{Barnault2003}
L.~Barnault and D.~Declercq, ``Fast decoding algorithm for {LDPC} over
  ${GF}(2^q)$,'' in \emph{The Proc. 2003 Inform. Theory Workshop}, 2003, pp.
  70--73.

\end{thebibliography}

\end{document}